\global\def\draftcontrol{0}

   \def\versionno{membrane2}

\catcode`\@=11

\expandafter\ifx\csname draftcontrol\endcsname\relax\global\def\draftcontrol{0}
\fi

{\count255=\time\divide\count255 by 60
\xdef\hourmin{\number\count255}
\multiply\count255 by-60\advance\count255 by\time
\xdef\hourmin{\hourmin:\ifnum\count255<10 0\fi\the\count255}}
\def\draftdate{\number\month/\number\day/\number\year\ \ \ \hourmin }

\newcommand\makepapertitle{\par
  \begingroup
    \renewcommand\thefootnote{\@fnsymbol\c@footnote}%
    \def\@makefnmark{\rlap{\@textsuperscript{\normalfont\@thefnmark}}}%
    \long\def\@makefntext##1{\parindent 1em\noindent
            \hb@xt@1.8em{%
                \hss\@textsuperscript{\normalfont\@thefnmark}}##1}%
     \newpage
     \global\@topnum\z@   
     \@makepapertitle
     \thispagestyle{empty}\@thanks
  \endgroup
  \setcounter{footnote}{0}%
  \global\let\thanks\relax
  \global\let\makepapertitle\relax
  \global\let\@makepapertitle\relax
  \global\let\@thanks\@empty
  \global\let\@author\@empty
  \global\let\@date\@empty
  \global\let\@title\@empty
  \global\let\title\relax
  \global\let\author\relax
  \global\let\date\relax
  \global\let\and\relax
  \def\version{\let\version\@version\@gobble}
}
\def\@makepapertitle{%
  \newpage
   \ifnum\draftcontrol=1 {}
   \version\versionno
   \vskip 3em%
   \else
   \hfill\hbox to 3cm {\parbox{4cm}{\@pubnum}\hss}%
   \vskip 3em%
   \fi
   \begin{center}%
   \let \footnote \thanks
     {\LARGE {\@title}}%
     \vskip 1.5em%
     {\normalsize
       \lineskip .5em%
       \begin{tabular}[t]{c}%
         \@author
       \end{tabular}\par}%
     \vskip 1.5em%
     {\@bstract}%
     \end{center}%
     \vskip 1.5em
     \@date%
   \par
}

\gdef\@pubnum{}
\def\pubnum#1{%
  \gdef\@pubnum{#1}}

\gdef\@bstract{}
\def\Abstract#1{%
  \gdef\@bstract{%
   \parbox{\textwidth-0pc}{%
   \centerline{\bf Abstract}\penalty1000%
\kern.2cm%
\noindent
\renewcommand\baselinestretch{1.0}%
{#1}}}
}

\def\ps@paper{\let\@mkboth\@gobbletwo%
     \ifnum\draftcontrol=1
    \def\@oddfoot{\hbox to \textwidth{\tiny \versionno \hfil\tiny\draftdate}%
    \hskip -\textwidth \hbox to \textwidth{\hfil\rm\thepage\hfil}}%
     \else\def\@oddfoot{\hbox to \textwidth{\hfil\rm\thepage\hfil}}
     \fi
     \let\@evenfoot\@oddfoot
}

\def\body{\clearpage
          \pagestyle{paper}
    }

\def\@version#1{\ifnum\draftcontrol=1
\typeout{}\typeout{#1}\typeout{}
\vskip3mm\centerline{\hbox{\fbox{\normalsize{\tt DRAFT -- #1 -- }
                   {\draftdate}}}}\vskip3mm
\fi}
\let\version\@version
\long\def\eqlabel#1{\ifnum\draftcontrol=1
                    \tag@false  
                    \tag*{(\theequation) \hbox to -0.2cm{\hspace{0cm}\small{#1}\hss}}
                    \refstepcounter{equation}
                    \edef\@currentlabel{\theequation}
                    \ltx@label{#1}          
                    \else
                    \label{#1}
                    \fi
                    }
\let\st@bibitem\@bibitem
\let\st@lbibitem\@lbibitem
\ifnum\draftcontrol=1
  \def\@bibitem#1{%
    \st@bibitem{#1}\a@@label{#1}\ignorespaces}
  \def\@lbibitem[#1]#2{%
    \st@lbibitem[#1]{#2}\a@@label{#2}\ignorespaces}
  \def\a@@label#1{%
    \gdef\a@lab{\smash{\normalfont\small#1}}
    \ifvmode
      \if@inlabel
        \global\setbox\@labels\hbox{%
          \llap{\a@lab\let\a@lab\relax
                \kern\@totalleftmargin\kern\marginparsep}%
          \box\@labels}%
      \fi
    \fi}
\fi

\documentclass[12pt,letterpaper]{article}

\usepackage{amsmath,amssymb,array,calc,epsfig,rotating,bm}
\usepackage[sort]{cite}
\usepackage{graphicx}
\usepackage{psfrag,verbatim}
\usepackage{xcolor}
\usepackage{hyperref}

\ifnum\draftcontrol=1
\fi

\renewcommand\baselinestretch{1.25}
\renewcommand\section{\@startsection {section}{1}{\z@}%
                                   {-3.5ex \@plus -1ex \@minus -.2ex}%
                                   {2.3ex \@plus.2ex}%
                                   {\normalfont\large\bfseries}}
\renewcommand\subsection{\@startsection{subsection}{2}{\z@}%
                                   {-3.25ex\@plus -1ex \@minus -.2ex}%
                                   {1.5ex \@plus .2ex}%
                                   {\normalfont\normalsize\bfseries}}
\renewcommand\subsubsection{\@startsection{subsubsection}{3}{\z@}%
                                   {-3.25ex\@plus -1ex \@minus -.2ex}%
                                   {1.5ex \@plus .2ex}%
                                   {\normalfont\normalsize\it}}
\renewcommand\paragraph{\@startsection{paragraph}{4}{\z@}%
                                   {-3.25ex\@plus -1ex \@minus -.2ex}%
                                   {1.5ex \@plus .2ex}%
                                   {\normalfont\normalsize\bf}}

\numberwithin{equation}{section}

\def\revise#1       {\raisebox{-0em}{\rule{3pt}{1em}}%
                     \marginpar{\raisebox{.5em}{\vrule width3pt\
                     \vrule width0pt height 0pt depth0.5em
                     \hbox to 0cm{\hspace{0cm}{%
                     \parbox[t]{4em}{\raggedright\footnotesize{#1}}}\hss}}}}

\newcommand\nxt[1]  {\\\fnxt#1}
\newcommand{\ie}{{\it i.e.,}\ }
\newcommand{\eg}{{\it e.g.,}\ }

\def\cala         {{\cal A}}

\def\calb         {{\cal B}}

\def\cald         {{\cal D}}

\def\calf         {{\cal F}}

\def\calk         {{\cal K}}

\def\calm         {{\cal M}}
\def\caln         {{\cal N}}
\def\calo         {{\cal O}}

\def\calv         {{\cal V}}

\def\complex      {{\mathbb C}}

\def\projective   {{\mathbb P}}

\def\reals        {{\mathbb R}}
\def\zet          {{\mathbb Z}}

\def\del          {\partial}

\def\Re           {{\rm Re\hskip0.1em}}
\def\Im           {{\rm Im\hskip0.1em}}

\def\sqr#1#2{{\vcenter{\vbox{\hrule height.#2pt
 \hbox{\vrule width.#2pt height#1pt \kern#1pt
 \vrule width.#2pt}\hrule height.#2pt}}}}

\newcommand{\kk}{\mathfrak{q}}
\newcommand{\ww}{\mathfrak{w}}

\def\aa1{\phi}
\def\cc1{\psi}

\catcode`\@=12

\begin{document}


\title{\bf On stability of baryonic black membranes}

\date{January 4, 2026}

\author{
Alex Buchel\\[0.4cm]
\it Department of Physics and Astronomy\\ 
\it University of Western Ontario\\
\it London, Ontario N6A 5B7, Canada
}

\Abstract{Near-extremal black membranes with topological (baryonic) $U(1)_B$
charge of M-theory compactified on the coset space $M^{1,1,0}$ are
stable.  $M^{1,1,0}$ coset is a $\zet_2$-invariant truncation of a
larger $Q^{1,1,1}$ coset, with diagonal $U(1)_B\equiv
U(1)_{B,+}\subset U(1)_B^2$ symmetry of the latter. We show that the
baryonic black membranes of M-theory $M^{1,1,0}$ compactifications are
unstable to $\zet_2$-odd gravitational bulk gauge and scalar
fluctuations, but only if this bulk scalar is identified with the
holographically dual $2+1$ dimensional superconformal gauge theory
operator of conformal dimension $\Delta=1$. The instability is
associated with the unstable charge transport of the off-diagonal
$U(1)_{B,-}\subset U(1)_B^2$ symmetry.
}

\makepapertitle
\bigskip

\body

\version\versionno
\tableofcontents

\section{Introduction and summary}\label{intro}
Near-extremal black branes with finite entropy density
in the limit of vanishing temperature $T\to 0$,
ubiquitous in the holographic correspondence
\cite{Maldacena:1997re,Aharony:1999ti}, recently
gained renewed interest as laboratories of quantum gravity
\cite{Turiaci:2023wrh}. The best explored holographic example
is that of the strongly coupled $\caln=4$ supersymmetric Yang-Mills (SYM) plasma
in four spacetime dimensions.
Here,  the equilibrium states of the gauge theory plasma,
with the same chemical potential $\mu$ for all $U(1)$ factors of the maximal
Abelian subgroup $U(1)^3\subset SU(4)$ $R$-symmetry, reach the quantum critical
regime as $\frac{T}{\mu}\to 0$. In the gravitational dual, such states are
represented by a Reissner--Nordstr\"{o}m (RN) black brane in asymptotically
$AdS_5$ spacetime. Unfortunately, in the extremal limit,
the black branes typically suffer from the the variety of instabilities:
the non-perturbative "Fermi-seasickness'' instability \cite{Hartnoll:2009ns},
the perturbative ``superconducting'' instability \cite{Hartnoll:2008vx},
or the perturbative ``charge-clumping''
instability  \cite{Gladden:2024ssb} --- either one of which precludes
reaching the interesting quantum critical regime. 

With the goal of constructing reliable (and stable) extremal
horizons, the authors of \cite{Herzog:2009gd,Klebanov:2010tj}
focused on holographic models   
from compactifications of string theory/M-theory
on $AdS_{p+2}\times Y$ manifolds with nonzero $p$th Betti number $b_p$, leading to $U(1)^{b_p}$ “baryonic” global symmetry.
Non-supersymmetric extremal quantum states supported by the
baryonic  $U(1)^{b_p}$ 
chemical potentials do not have superconducting instabilities. As an example,
consider strongly coupled $\caln=1$ $SU(N)\times SU(N)$ gauge theory in four
spacetime dimensions, the Klebanov--Witten (KW) model \cite{Klebanov:1998hh}.
The theory has $U(1)_R\times U(1)_B$ global symmetry, which supports
quantum critical states charged under either of the $U(1)$s.
The $R$-symmetry charged quantum critical states are unstable 
due to the condensation of the chiral primary
$\calo_F\equiv {\rm Tr} (W_1^2+W_2^2)$, where $W_i$ are the gauge superfields corresponding
to the two gauge group factors of $SU(N)\times SU(N)$ quiver \cite{Buchel:2024phy}. 
The gauge-invariant operators  of the KW theory charged under $U(1)_B$ have conformal
dimensions of order\footnote{The smallest such operators involve determinants of the
bifundamental matter fields of the KW quiver gauge theory. This justifies the nomenclature “baryonic symmetry”.} $N$, with the charge-to-mass ratio
too small to trigger the superconducting instability \cite{Herzog:2009gd}. 
Nonetheless, quantum critical states with a baryonic charge of the KW theory are
unstable \cite{Buchel:2025jup}:
even though such states have zero $R$-symmetry charge density, at low temperatures
$R$-charge starts
“clumping”, breaking the homogeneity of $U(1)_B$ charged
thermal equilibrium state.\footnote{This is a direct consequence of the
thermodynamic instabilities of the underlying thermal states \cite{Buchel:2005nt}. For charged plasma this was originally explained in
\cite{Gladden:2024ssb,Gladden:2025glw}.
}

So far, the only known example of the stable non-supersymmetric
extremal horizon of string theory/M-theory  
 is realized in a membrane theory of Klebanov, Pufu and
Tesileanu (KPT) \cite{Klebanov:2010tj}. The KPT model
is a holographic example of a three dimensional superconformal
gauge theory arising from compactification of M-theory
on regular seven-dimensional Sasaki--Einstein manifold --- 
$\frac{SU(3)\times SU(2)}{SU(2)\times U(1)}$ coset, known as
$M^{1,1,0}$. Much like the KW theory, the
holographic membrane model of M-theory on $M^{1,1,0}$ has
$U(1)_R\times U(1)_B$ global symmetry.
Here, there are three distinct
near-extremal regimes: one supported by the $U(1)_R$ charge density,
and the other two supported by the $U(1)_B$ charge density.
The reason for the distinct baryonic near-criticality comes from the
fact that the dual gravitational backgrounds
have nontrivial support from the bulk scalar with $m^2 L^2=-2$,
corresponding to an operator of conformal dimension $\Delta=(2,1)$.
Depending on whether one uses a normal or an alternative quantization \cite{Klebanov:1999tb}, one obtains either of two field theory duals,
each with a near-extremal regime.  It was shown in \cite{Buchel:2025ven}
that only the baryonic black membranes are stable:
extremal horizons supported by $U(1)_R$ charge density
suffer from both the (threshold) superconducting and
$U(1)_B$ charge clumping instabilities\footnote{The $U(1)_B$
charge transport instability sets in at higher temperatures.}.

In this paper we further explore instabilities of the
KPT baryonic membranes. There is a larger consistent
truncation of M-theory on $\frac{SU(2)^3}{U(1)^2}$ coset, known as $Q^{1,1,1}$,
which is a $U(1)$ fibration over
$\complex\projective^1\times\complex\projective^1\times\complex\projective^1$. This manifold has the second Betti number $b_2=2$, so that the
corresponding boundary superconformal theory 
has $U(1)_R\times U(1)_B^2$ global symmetry.
The KPT model  is a consistent truncation of  $Q^{1,1,1}$ M-theory
compactification where the two Betti vector multiples
of  $Q^{1,1,1}$ are identified. From this perspective the KPT model
is a $\zet_2$-even sector
of $Q^{1,1,1}$ model under the interchange of the Betti multiples,
and its  baryonic symmetry $U(1)_B\equiv U_{B,+}$ is a diagonal
subgroup of $U(1)_B^2$ of this
larger model.  We present detailed analysis of the hydrodynamic
transport of the off-diagonal $U(1)_{B,-}\subset U(1)_B^2$
charge density fluctuations.
This $\zet_2$-odd sector of the gravitational dual fluctuations
includes a massless Betti vector $A_-$ and a scalar $v_-$
with $m^2 L^2 =-2$, corresponding to an operator $\calo_-$ of conformal
dimension $\Delta=(2,1)$. Here, once again we can use either a normal
$\dim{\calo_-}=2$, or an alternative  $\dim{\calo_-}=1$ quantization. 
We find that the diffusion coefficient
$D_{B,-}$ of the 
charge density fluctuations associated with $U(1)_{B,-}$
symmetry becomes negative below some critical
temperature $T_{crit}$, relative to the $U(1)_{B,+}$-charge chemical potential
$\mu_B$ of the near-critical thermal equilibrium states of KPT plasma,
provided $\dim{\calo_-}=1$; $D_{B,-}>0$ at any temperature if
$\dim{\calo_-}=2$:  
\begin{equation}
\begin{cases}
D_{B,-}>0\,,\qquad {\frac{T}{\mu_B}> \frac{T}{\mu_B}\bigg|_{crit}}\,,\qquad \dim{\calo_-}=1\,,\\
D_{B,-}<0\,,\qquad {\frac{T}{\mu_B}< \frac{T}{\mu_B}\bigg|_{crit}}\,,\qquad \dim{\calo_-}=1\,,\\
D_{B,-}>0\,,\qquad \frac{T}{\mu_B}\ {\rm is\ any}\,,\qquad \dim{\calo_-}=2\,.
\end{cases}
\eqlabel{dinb}
\end{equation}
Whenever $D_{B,-}<0$, the $U(1)_{B,-}$ charge density
is unstable to clumping: indeed,
in the hydrodynamic\footnote{We use notations
$\ww\equiv\frac{w}{2\pi T}$ and $\kk\equiv \frac{|\vec k|}{2\pi T}$ where $e^{-i w t + i \vec k \cdot \vec x}$ is the profile of the hydrodynamic perturbation.}  $\kk\to 0$  limit, the dispersion relation
characterizing the charge diffusion is 
\begin{equation}
\ww=-i D_{B,-}\kk^2+\calo(\kk^2)\,,
\eqlabel{diffinb}
\end{equation}
thus
\begin{equation}
D_{B,-}<0\qquad \Longleftrightarrow\qquad \Im[\ww]>0\,.
\eqlabel{inst}
\end{equation}
The precise value of the critical temperature in \eqref{dinb}
depends on the quantization of the $m^2L^2=-2$ scalar supporting
the background geometry.

The rest of the paper is organized as follows.
In the next section we summarize\footnote{See \cite{Buchel:2025ven}
for additional details.} the relevant effective action
for the M-theory flux compactifications on  $Q^{1,1,1}$.
We review the background geometry describing
baryonic black branes. In section \ref{fluctuations}
we compute $D_{B,-}$ for different quantizations
of the background scalar supporting
the near-extremal baryonic black branes, and
for different quantizations of the $\zet_2$-odd bulk scalar
$v_-$. Additionally, in section \ref{hom}, we argue that there are no
homogeneous and isotropic equilibrium phases of the
baryonic black branes with spontaneously broken $\zet_2$
symmetry. Appendices collect technical details
necessary to reproduce the claims of the paper.

The main conceptual point of the paper is that the onset of
instabilities in the extremal limit of the boundary CFTs, with the
same global symmetries and the conserved currents, depends on the
details of their operator spectra, \ie on the identification of the
bulk scalar $v_-$ with the operator $\calo_-$ either of the conformal
dimension $\dim \calo_-\equiv \dim\calo_-^{(-)}=1$ or
$\dim \calo_-\equiv \dim\calo_-^{(+)}=2$. The CFT with a $\calo_-^{(-)}$
operator is unstable in the extremal limit, while a CFT with a
$\calo_-^{(+)}$ operator is stable in the extremal limit.
Interestingly\footnote{I am grateful to a Referee for raising this
issue.}, both CFTs are related by a holographic RG flow, triggered by
a relevant double-trace deformation
\cite{Aharony:2001pa,Witten:2001ua,Berkooz:2002ug,Gubser:2002vv}
$\int f \left(\calo_{-}^{(-)}\right)^2$.  Thus, there is a potential
for a new critical phenomenon, where the onset of the dynamical
instability depends on how large the double-trace mass scale $f$ is
compare to the $U(1)_{B,+}$ baryonic chemical potential $\mu_B$.  We
will present such analysis elsewhere.

We leave open for the future work the study of the
end point of the identified instability in the CFT
with a $\calo_-^{(-)}$ operator in the spectrum: while there
are multiple occurrences of the related
instabilities in other models
\cite{Gladden:2024ssb,Buchel:2025jup,Gladden:2025glw,Buchel:2025ven},
their ultimate fate is unknown (and unfortunately, likely to
be difficult to analyze). There is also currently no
understanding what underlying physics makes some models stable
in the extremal limit, and the other ones not.

\section{M-theory on $Q^{1,1,1}$ and near-extremal baryonic black branes}
\label{action}

Effective four-dimensional action of
$\caln=2$ gauged supergravity describing flux compactifications of M-theory
on $Q^{1,1,1}$ is given by \cite{Cassani:2012pj,Buchel:2025ven}
\begin{equation}
\begin{split}
&S_{Q^{1,1,1}}=\frac {1}{\kappa_4^2}\int_{\calm_4}
\biggl[
\frac 12 R_4\star 1-\biggl\{\ (\del\phi)^2+g_{ij} \del t^i\del \bar{t}^j
\ \biggr\}
\star 1 
-\frac 14 e^{-4\phi} dB\wedge\star dB\\
&\qquad +\frac 14\Im
\caln_{IJ} F^I\wedge\star F^J+\frac 14\Re\caln_{IJ} F^I\wedge F^J-\frac 12 e_0\ dB\wedge A^0
-V_{Q^{1,1,1}}\star 1\biggr]\,,\\
&V_{Q^{1,1,1}}=e^{4\phi} \calk\cdot\sum_i v_i^{-2}-8
e^{2\phi}\cdot\sum_i v_i^{-1}+\frac{e^{4\phi}}{4} \calk^{-1}\cdot\
\sum_{k} \biggl[\sum_{ij}\ \calk_{ijk}\ b_im_j v_k\
\biggr]^2
\\&\qquad +\frac {e^{4\phi}}{4} \calk^{-1}\cdot
\biggl[
e_0+\frac 12 \sum_{i,j,k} \calk_{ijk}\ b_ib_jm_k\biggr]^2\,,
\end{split}
\eqlabel{actionckv}
\end{equation}
with $t^i\equiv v_i+i b_i$. Here:
\begin{itemize}
\item $\{I,J\}=\{0,1,2,3\}$, $\{i,j,k\}=\{1,2,3\}$, and $m^I=\{0,2,2,2\}$.
The constant $e_0$ sets the radius $L$ of the asymptotic $AdS_4$ spacetime;
in what follows we will choose $e_0=6\Longrightarrow L=\frac 12$.
\item $B$ is a 2-form on $\calm_4$; $A^I$ are the 1-forms on $\calm_4$
with the
field strength $\calf^I\equiv dA^I$, and the generalized field strengths
$F^I$ are defined as $F^I=\calf^I-m^I B$. $t^i$ and $\phi$
are 0-forms on $\calm_4$.
\item Explicit expression for the gauge kinetic matrix is
\begin{equation}
\begin{split}
&\Re \caln_{00}=-\frac 13 \calk_{ijk}b_ib_jb_k\,,\qquad \Re \caln_{0i}=\frac12
\calk_{ijk} b_jb_k\,,\qquad \Re\caln_{ij}=-\calk_{ijk}b_k\,,\\
&\Im \caln_{00}=-\calk(1+4 g_{ij}b_ib_j)\,,\qquad \Im \caln_{0i}=4\calk g_{ij}b_j\,,\qquad \Im\caln_{ij}=-4\calk g_{ij}\,,
\end{split}
\eqlabel{flux}
\end{equation}
where $\calk_{ijk}=1$ for $i\ne j\ne k$ and $0$ otherwise,
$\calk=\prod_i v_i$, and $g_{ij}=\frac 14
v_i^{-2}\ \delta_{ij}$.
\end{itemize}

Consistent sub-truncation of the effective action
\eqref{actionckv} $S_{Q^{1,1,1}}\to S_{M^{1,1,0}}$ identifies
the Betti vector multiples
\begin{equation}
\{A^3,t^3\}\equiv \{A^1,t^1\}\,.
\eqlabel{mtruncation}
\end{equation}
Near-extremal black membranes with a $U(1)_B$ chemical potential
are homogeneous and isotropic solutions of the effective action
 $S_{M^{1,1,0}}$ with \cite{Klebanov:2010tj,Buchel:2025ven}
 \begin{equation}
 A^0\equiv 0\,,\qquad b_i\equiv 0\,,\qquad B\equiv 0\,,
 \eqlabel{truncated}
 \end{equation}
the background 4D metric on $\calm_4$ and the remaining 2-form field strengths
$\{\calf^1,\calf^2\}$ as 
\begin{equation}
ds_4^2=-\frac{4\alpha^2 f}{r^2}\ dt^2+\frac{4\alpha^2}{r^2} d\bm{x}^2
+ \frac{s^2}{4r^2 f}\ dr^2\,,\ \ \calf^1=\frac{q\alpha s}{v_2}\ dr\wedge dt
\,,\ \ \calf^2=-\frac{2v_2^2}{v_1^2}\calf^1\,,
\eqlabel{4dm}
\end{equation}
where $\alpha,q$ are constant coefficients, and $f,s,v_1,v_2,g\equiv e^\phi$ are all functions of
the radial coordinate
\begin{equation}
r\in (0,1)\,.
\eqlabel{defr}
\end{equation}
The asymptotic $AdS_4$ boundary is located as $r\to 0_+$, requiring
\begin{equation}
\lim_{r\to 0_+} \{f,s,v_1,v_2,g\}(r)=1\,,
\eqlabel{uvboundary}
\end{equation}
and the regular Schwarzschild horizon is located at a simple root
of the blackening factor $f$, with all the other bulk fields being finite.
Using a constant rescaling of a radial coordinate $r\to \lambda r$
we can always assume that
the horizon is as $r\to 1_-$, thus requiring
\begin{equation}
\lim_{x\to 1_- } f(r)=0\,,\qquad \lim_{r\to 1_-} \{s,v_1,v_2,g\}(r)={\rm finite}\,.
\eqlabel{irboundary}
\end{equation}
The constant $\alpha$  in \eqref{4dm}
is a scale resulting
from fixing the horizon location as in \eqref{irboundary};
it is
necessary to define the temperature $T\propto |\alpha|$ and the
chemical potential
$\mu_B\propto \alpha$ of the boundary superconformal theory thermal state,
dual to a baryonic black membrane geometry \eqref{4dm}.
This constant will drop out from all the dimensionless thermodynamic 
ratios, \eg $\frac {T}{\mu_B}$.
The dimensionless parameter $q$ in \eqref{4dm}
is related to a baryonic chemical potential:
specifically, the conserved $U(1)_B$ current of the boundary
2+1 dimensional superconformal gauge theory is  holographically
dual to a bulk 1-form gauge potential $A^1$,
\begin{equation}
dA^1=\calf^1=\frac{q\alpha s}{v_2}\ dr\wedge dt\qquad \Longrightarrow\qquad
\frac{d A^1_t}{dr}=\frac{q\alpha s}{v_2}\,,
\eqlabel{defa1}
\end{equation}
thus we require
\begin{equation}
\lim_{r\to 0_+} A^1_t(r)=\mu_B\,,\qquad \lim_{r\to 1_-} A^1_t(r)=0\,.
\eqlabel{a1tboundary}
\end{equation}
The holographic spectroscopy
relates the bulk scalars $\{v_1,v_2,g\}$ to the boundary gauge theory operators
$\calo_\Delta$ of conformal dimension $\Delta$ as in table \ref{table1}:

\begin{table}[ht]
\caption{Holographic spectroscopy of the bulk scalars
supporting baryonic black membranes \cite{Cassani:2012pj}}
\centering
\begin{tabular}{c c c }
\hline
mass eigenstate & $m^2 L^2$ & $\Delta$ \\
\hline
$\ln [v_1v_2^{-1}]$ & $-2$ & $(2,1)$  \\
\hline
$\ln [v_1^2v_2 g^3]$ & $4$& $4$\\
\hline
$\ln [v_1^2v_2 g^{-4}]$ & $18$& $6$ \\
\hline
\end{tabular}
\label{table1}
\end{table}

The bulk scalar $\ln[v_1 v_2^{-1}]$ can be
identified 
either with the operator $\calo_2$, the {\it normal quantization},
or with the operator $\calo_1$, the {\it alternative quantization}.
Each of the identifications allows for a nonsingular extremal limit
of the baryonic black membrane \eqref{4dm} $\calm_4\to AdS_2\times \reals^2$,
\ie the limit of vanishing of its Hawking temperature $T\to 0$.
Notice that at extremality the Bekenstein entropy density
$s$ of the membrane remains finite,
\begin{equation}
s=\frac{2\pi}{\kappa_4^2}\ \lim_{r\to 1_-} \frac{4\alpha^2}{r^2} =
\frac{8\pi\alpha^2}{\kappa_4^2}\,,
\eqlabel{defs}
\end{equation}
while the dimensionless $\alpha$-independent ratio $\frac{s}{T^2}\to \infty $.

The equations of motion for the baryonic black membrane
background fields $\{f,s,v_1,v_2,g\}$ derived from the
effective action $S_{M^{1,1,0}}$ are collected in
appendix \ref{background}, along with the
near-boundary $r\to 0_+$ and the near-horizon $r\to 1_-$ asymptotic
expansions, enforcing the boundary conditions \eqref{uvboundary}
and \eqref{irboundary}. Explicit expressions for $T$ and
$\mu_B$ are given by \eqref{tbaryonic} and \eqref{mubaryonic}.
As the baryonic black membrane temperature varies as
$\frac{T}{\mu_B}\in (0,\infty)$, 
parameter $q\in(0,q_{crit}=2^{15/4}/3^{5/4})$, with \cite{Buchel:2025ven}
\begin{equation}
\lim_{q\to 0}\ \frac{T}{\mu_B} =\infty\,,\qquad \lim_{q\to q_{crit}}\
\frac{T}{\mu_B} \propto \left(1-\frac{q}{q_{crit}}\right)\to 0\,.
\eqlabel{limq}
\end{equation}

\section{$\zet_2$-odd fluctuations and the $U(1)_{B,-}$ charge transport}
\label{fluctuations}

In this section we consider fluctuations about the baryonic
black membrane solution \eqref{truncated}, \eqref{4dm} that are odd
with respect to the interchange of the Betti vector multiples
of the effective action $S_{Q^{1,1,1}}$,
\begin{equation}
\{A^3,t^3\}\qquad \longleftrightarrow\qquad \{A^1,t^1\}\,.
\eqlabel{switch}
\end{equation}
Specifically, we introduce linearized fluctuation $\{\delta v_-, \delta b_-,
\delta A_-\}$ as 
\begin{equation}
\begin{split}
&t_1=v_1 e^{\frac 12 \delta v_-}+i \frac12\delta b_-\,,\qquad
t_3= v_1^{-\frac 12 \delta v_-}-i\frac 12\delta b_-\,,\\
&A^1\to A^1 +\frac 12 \delta A_-\,,\qquad A^3\to A^1-\frac 12 \delta A_-\,,
\end{split}
\eqlabel{parm}
\end{equation}
so that under \eqref{switch},
\begin{equation}
\{\delta v_-, \delta b_-,
\delta A_-\}\qquad \longrightarrow\qquad -\ \{\delta v_-, \delta b_-,
\delta A_-\}\,.
\eqlabel{odd}
\end{equation}
As the fluctuations of all the other fields of the
baryonic black membranes within $S_{Q^{1,1,1}}$ are
even\footnote{These modes were studied in details in
\cite{Buchel:2025ven}.} under \eqref{switch}, $\zet_2$-odd
modes will decouple, governed by the quadratic action
$S_-\{\delta v_-, \delta b_-,
\delta A_-\}\equiv S_{Q^{1,1,1}}-S_{M^{1,1,0}}$,
\begin{equation}
\begin{split}
&S_{-}=\frac{1}{\kappa_4^2}\int_{\calm_4} \biggl[-\biggl\{
\frac{1}{8v_1^2}(\del \delta b_-)^2+\frac 18(\del\delta v_-)^2
\biggr\}\star 1-\frac {v_2}{8}\ \delta \calf_-\wedge\star \delta \calf_--
\frac{v_2}{4} (\delta v_-)^2 \calf^1\wedge \star \calf^1\\
&+\frac{v_2}{2} \delta v_-\ \delta\calf_-\wedge \star \calf^1
+\frac 14 \delta b_-\ \delta\calf_-\wedge \calf^2-V_-\star 1 
\biggr]\,,\\
&V_-=\frac{g^2 (g^2v_1v_2-2)}{v_1}\ (\delta v_-)^2
+\frac{g^4(v_1^2-3)}{2v_2v_1^2}\ (\delta b_-)^2\,,\qquad
\delta\calf_-\equiv dA_-\,.
\end{split}
\eqlabel{sm}
\end{equation}
Within the effective action \eqref{sm}, we further consider
 fluctuations  to be functions of $t$, $x_2$, and $r$ as follows 
\begin{align}
\delta A_- &=e^{-i w t +i k x_2}\ \biggl
( \cala_t\ dt+ \cala_{2}\ d x_2+ \cala_r\ dr\biggr)\,,\quad
\{\delta v_-\,,\,\delta b_-\}=e^{-i w t +i k x_2}\ \{\calv\,,\,\calb\}\,,
\eqlabel{fldec}
\end{align}
where $\{\cala_{t,2,r},\calv,\calb\}$ are functions of the radial coordinate $r$. 
We use the bulk gauge transformations of Betti vectors $A^1$ and $A^3$ to
set\footnote{This would lead to the first-order constraint \eqref{flm1}.}
\begin{equation}
\cala_r=0\,.
\eqlabel{gauge}
\end{equation}
The equations of motion for the fluctuations are collected in
appendix \ref{flucm}.
Following \cite{Cassani:2012pj}, the holographic spectroscopy relates
the (pseudo-)scalar modes $\{\calv,\calb\}$ to the boundary gauge
theory operators
$\delta\calo^\calv_\Delta,\delta\calo^\calb_\Delta,$ of conformal dimension $\Delta$
as in table \ref{table2}.
\begin{table}[ht]
\caption{Holographic spectroscopy of $\zet_2$-odd (pseudo-)scalars}
\centering
\begin{tabular}{c c c}
\hline
mass eigenstate & $m^2 L^2$ & $\Delta$ \\
\hline
$b_1-b_3$ & $-2$ & $(2,1)$  \\
\hline
$\ln[v_1v_3^{-1}]$ & $-2$& $(2,1)$\\
\hline
\end{tabular}
\label{table2}
\end{table}
Here again, we have the choice to quantize the fluctuations so that they correspond either to  boundary CFT operators of dimension 2 (the \textit{normal quantization}), or to operators of
dimension 1  (the \textit{alternative quantization}).
This choice is independent from the choice of quantization
for the background solution. 

We find that the fluctuations $\{\cala_t,\cala_2,\calv\}$
decouple from $\calb$ --- the former describe the $U(1)_{B,-}$
charge transport, the while latter can lead to potential threshold instabilities
(to be further discussed in section \ref{hom}). To proceed with the $U(1)_{B,-}$
charge transport we introduce
\begin{equation}
Z\equiv \kk\ \cala_t + \ww\ \cala_2\,.
\eqlabel{defz}
\end{equation}
We use the constraint \eqref{flm1}  to obtain from \eqref{flm2}-\eqref{flm4}
a decoupled
set of the second-order equations for
\begin{equation}
\{\ Z\,,\ \calv\ \}\,.
\eqlabel{set}
\end{equation}
Solutions of the resulting equations with appropriate boundary conditions
determine the spectrum of $U(1)_{B,-}$ charged quasinormal 
modes of the baryonic black membranes  --- equivalently
the $\zet_2$-odd subsector of physical spectrum of linearized fluctuations in membrane gauge theory plasma with
a baryonic chemical potential. Following \cite{Kovtun:2005ev,Son:2002sd}
we impose the incoming-wave boundary conditions at the black membrane
horizon, and 'normalizability'
at asymptotic $AdS_4$ boundary.
Focusing on the $\Re[\ww]=0$  diffusive branch, and
introducing
\begin{equation}
\begin{split}
&Z=\left(1-r\right)^{-i\ww/2}\ z\,,\qquad \calv=(1-r)^{-i\ww/2}\ u\,,\qquad
\ww=-i v\ \kk\,,
\end{split}
\eqlabel{income}
\end{equation}
we solve the fluctuation equations  subject to the asymptotics: 
\nxt in the UV, \ie as $r\to 0_+$, and with the
identifications\footnote{Likewise, we develop the UV expansions for the
alternative quantization of either the background, $\ln[v_1 v_2^{-1}]$, or the fluctuation  scalar, $u$: $\{\calo_2,\delta\calo^\calv_1\}$,
$\{\calo_1,\delta\calo^\calv_2\}$, and $\{\calo_1,\delta\calo^\calv_1\}$.}
$\ln [v_1v_2^{-1}]\Longleftrightarrow\calo_2$ and $u
\Longleftrightarrow\delta\calo_2^\calv$, 
\begin{equation}
\begin{split}
&z=\kk\ r -\frac 12 \kk^2 v\ r^2+\calo(r^3)\,,\qquad
u=u_2\ r^2-\frac12\kk v u_2\ r^3+\calo(r^4)\,,
\end{split}
\eqlabel{uv1}
\end{equation}
specified, for a fixed background and a momentum $\kk$,  by
\begin{equation}
\biggl\{\
v\,,\ u_2
\
\biggr\}\,;
\eqlabel{fluv}
\end{equation}
\nxt in the IR, \ie as $y\equiv 1-r\to 0_+$,
\begin{equation}
\begin{split}
&z=z^h_0+\calo(y)\,,\qquad 
u=u^h_{0}+\calo(y)\,,
\end{split}
\eqlabel{ir1}
\end{equation}
specified  by
\begin{equation}
\biggl\{\
z^h_0\,,\ u^{h}_{0} 
\
\biggr\}\,.
\eqlabel{flir}
\end{equation}
Note that in total we have $2+2=4$ parameters, see \eqref{fluv} and \eqref{flir},
which is precisely what is necessary to identify a solution of a coupled system of
2 second-order ODEs for $\{z, u\}$. Furthermore, since the equations are linear in the
fluctuations, we can, without loss of generality, normalize the solutions so that
\begin{equation}
\lim_{r\to 0}\ \frac{dz}{dr}=\kk\,.
\eqlabel{normaliz}
\end{equation}

Once we fix the background, and solve the fluctuation equations of motion,
we obtain $v=v(\kk)$. Given $v$ we extract the $U(1)_{B,-}$-charge
diffusion coefficient $\cald_{B_-}$, as 
\begin{equation}
\ww =-i\cdot\ \underbrace{2\pi T D_{B,-}}_{\equiv \cald_{B,-}}\cdot\ \kk^2+\calo(\kk^3)\,,\qquad 
\cald_{B,-}\equiv \frac{dv}{d\kk}\bigg|_{\kk=0}\,.
\eqlabel{defd}
\end{equation}

For general values of $q$ we have to solve the fluctuation equations
numerically. At $q=0$, an analytic solution is possible in the
limit $\kk\to 0$ --- which is precisely what is needed to extract $\cald_{B,-}$
\cite{Buchel:2025ven}:
\begin{equation}
\cald_{B,-}\bigg|_{q=0}=\frac 32\,.
\eqlabel{an3}
\end{equation}

\begin{figure}[t]
\begin{center}
\psfrag{d}{{$\cald_{B,-}$}}
\psfrag{q}{{$q/q_{crit}$}}
  \includegraphics[width=4in]{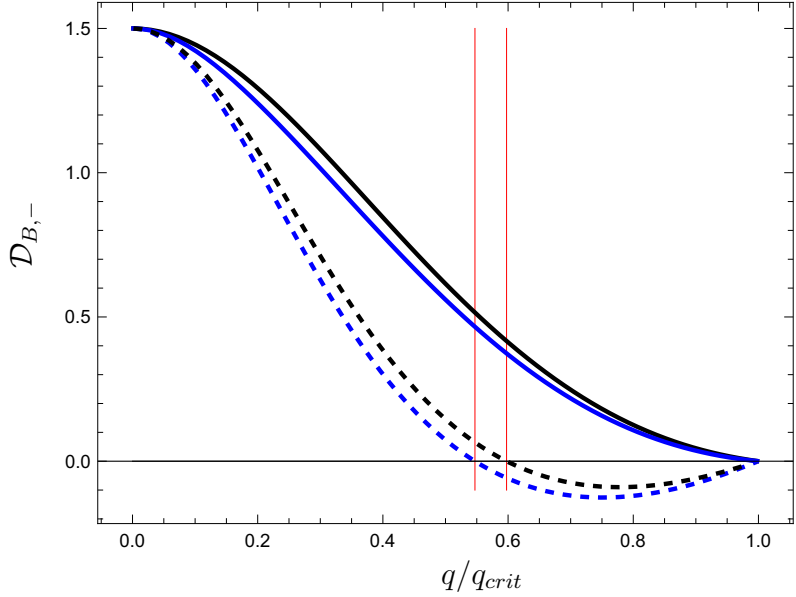}
\end{center}
 \caption{$U(1)_{B,-}$-charge dimensionless diffusion coefficient
 $\cald_{B,-}=2\pi T D_{B,-}$ of the baryonic membrane theory plasma
 for different quantizations
 of the gravitational dual
 scalars $\{\ln[v_1v_2^{-1}],\delta
 \ln[v_1v_3^{-1}]\}$: $\{\calo_2,\delta\calo_2^\calv\}$ (black,solid),
  $\{\calo_2,\delta\calo^\calv_1\}$ (black,dashed),
  $\{\calo_1,\delta\calo_2^\calv\}$ (blue,solid),
  $\{\calo_1,\delta\calo^\calv_1\}$ (blue,dashed).
At $q=0$, $\cald_{B,-}=\frac 32$ \eqref{an3}, while it vanishes in
the quantum critical regime $q\to q_{crit}$,  $\cald_{B,-}\propto \frac{T}{\mu_B}\to 0$.
Independent of the background scalar $\ln[v_1v_2^{-1}]$ quantization,
there is an onset of the $U(1)_{B,-}$ charge clumping instability  
for $\delta \ln[v_1v_3^{-1}]\Longleftrightarrow \delta\calo_1^\calv$ quantization
(the dashed curves), represented by vertical red lines.
}\label{figure1}
\end{figure}

For $q\in(0,q_{crit})$  the $U(1)_{B,-}$-charge
diffusion coefficient of the baryonic membrane theory plasma
is computed numerically, see fig.~\ref{figure1}.
Black curves correspond to the background scalar quantization
as $\ln[v_1v_2^{-1}]\Longleftrightarrow \calo_2$, while the
blue curves correspond to the quantization $\ln[v_1v_2^{-1}]\Longleftrightarrow \calo_1$.
Furthermore, the solid curves represent $\zet_2$-odd scalar $\ln[v_1v_3^{-1}]$ quantization
as  $\delta \ln[v_1v_3^{-1}]\Longleftrightarrow \delta\calo_2^\calv$, while the dashed
curves represent $\delta \ln[v_1v_3^{-1}]\Longleftrightarrow \delta\calo_1^\calv$.
In the latter case (the dashed curves), there is  the $U(1)_{B,-}$ 
charge clumping instability for $q>q_{unstable}$
(correspondingly  $\frac{T}{\mu_B}<\frac{T}{\mu_B}\bigg|_{crit}$), represented by vertical red lines,
\begin{equation}
\begin{array}{c|c|c|c} \{\calo,\delta\calo^\calv\}
\qquad & {\rm fig.~\ref{figure1}\ curve\ style}& {q_{unstable}}/{q_{crit}}\qquad & {T}/{\mu_B}|_{crit}\qquad\\ \hline
\{\calo_2,\delta\calo_1^\calv\} & {\rm (black,dashed)} & 0.597(6) &  0.251(5)\\ \{\calo_1,\delta\calo_1^\calv\}
& {\rm (blue,dashed)} & 0.547(1) &  0.276(2)\end{array}
\eqlabel{master}
\end{equation}
To compute $\frac{T}{\mu_B}\bigg|_{crit}$ for a given value of $\frac{q_{unstable}}{q_{crit}}$
we use \eqref{tbaryonic} and \eqref{mubaryonic}.

\section{Threshold instabilities from condensation of
$\{\delta v_-, \delta b_-\}$ }\label{hom}

Consider spatially homogeneous and isotropic fluctuations of the bulk
(pseudo-)scalars $\calv$ and $\calb$ about baryonic black membrane of section \ref{action}.
The corresponding equations of motion can be obtained from
\eqref{flm1}-\eqref{flm5} in the limit
\begin{equation}
\{w,k\}\to 0\,,
\eqlabel{homlimb}
\end{equation}
provided we set $\cala_2=0$. We find two decoupled sets:
\begin{itemize}
\item $\{\calv,a\equiv \cala_t'\}$,
\begin{equation}
\begin{split}
&0=a'-\frac{2 q s}{v_2} \calv'+\biggl(
\frac{(v_2')^2 r}{v_2^2}+\frac{2 (v_1')^2 r}{v_1^2}+\frac{4 (g')^2 r}{g^2}
+\frac{4 v_2'}{v_2}
\biggr) \frac a4\,,
\end{split}
\eqlabel{homs11}
\end{equation}
\begin{equation}
\begin{split}
&0=\calv''+\biggl(
\frac{s^2 (v_1^4+2 v_1^2 v_2^2+9)}{4r f v_2 v_1^2} g^4
-\frac{2 s^2 (v_1+2 v_2)}{v_1 r f v_2} g^2+\frac{s^2 q^2 r^3 (v_1^2+2 v_2^2)}{8f v_2 v_1^2} 
+\frac1r\biggr) \calv'
\\&+\biggl(
-\frac{2 s^2 v_2}{f r^2} g^4+\frac{4 s^2}{v_1 f r^2} g^2+\frac{s^2 q^2 r^2}{2f v_2}
\biggr) \calv-\frac{s a q r^2}{2f}\,;
\end{split}
\eqlabel{homs12}
\end{equation}
\item $\{\calb\}$,
\begin{equation}
\begin{split}
&0=\calb''+\biggl(
\frac{s^2 (v_1^4+2 v_1^2 v_2^2+9)}{4r f v_2 v_1^2} g^4
-\frac{2 s^2 (v_1+2 v_2)}{v_1 r f v_2} g^2
+\frac{s^2 q^2r^3 (v_1^2+2 v_2^2)}{8f v_2 v_1^2} -\frac{2 v_1'}{v_1}\\&+\frac 1r
\biggr) \calb'-\frac{s^2 \calb g^4 (v_1^2-3)}{v_2 f r^2}\,.
\end{split}
\eqlabel{homs2}
\end{equation}
\end{itemize}
$\ $
\nxt In the UV, \ie as $r\to 0_+$, and with the identification\footnote{Likewise, we develop the UV expansions for the alternative quantization of the
background scalar $\ln[v_1 v_2^{-1}]\Longleftrightarrow\calo_1$.}
$\ln[v_1 v_2^{-1}]\Longleftrightarrow\calo_2$, 
\begin{equation}
\begin{split}
&a=2 q u_1\ r+2 q u_2\ r^2+\calo(r^3)\,,\qquad \calv=u_1\ r +u_2\ r^2
+\calo(r^4)\,,\\
&\calb=\calb_1\ r +\calb_2\ r^2+\calo(r^4)\,.
\end{split}
\eqlabel{homuv1}
\end{equation}
Notice that $\lim_{r\to 0}a=0$ --- this ensures that the fluctuations
$\{\calv,a\}$ have the vanishing $U(1)_{B,-}$ charge.  
In the quantization where $\calv$ (or $\calb$) is identified with the boundary
gauge theory operator $\delta\calo_2^\calv$ (correspondingly $\delta\calo_2^\calb$)
the coefficient $u_1$
(correspondingly $\calb_1$)
is the source, while in the identification
$\calv\Longleftrightarrow\delta\calo^\calv_1$ (or $\calb\Longleftrightarrow\delta\calo^\calb_1$)
the source term is $u_2$ (correspondingly $\calb_2$).
\nxt In the IR, \ie as $y\equiv 1-r\to 0$, 
\begin{equation}
\begin{split}
&\calv=u_{0}^h+\calo(y)\,,\qquad a=a_{0}^h+\calo(y)\,,\\
&\calb=\calb_{0}^h+\calo(y)\,.
\end{split}
\eqlabel{b1b2hb}
\end{equation}
Following \cite{Buchel:2019pjb}, \eg  to identify the onset of the instability
associated with the condensation of $\delta\calo_2^\calv$ 
we keep fixed the source term of the operator, $u_{1}=1$,
and scan $q$ (correspondingly $\frac{T}{\mu_B}$), 
looking for a divergence of the expectation value of the corresponding
operator $\langle\delta\calo_2^\calv\rangle\propto u_2$.
A divergence signals the presence of a homogeneous and isotropic
normalizable mode of the fluctuations of $\calv$ --- the threshold
for the instability. 
We performed all such scans, for both quantizations of the background
scalar $\ln[v_1v_2^{-1}]$, and independently for both quantizations of the $\zet_2$-odd
(pseudo-)scalars $\calv, \calb$ ---  there are no divergences of the expectation values
of the corresponding operators.

\section*{Acknowledgments} 
This work was supported by NSERC through the Discovery Grants program.

\appendix

\section{Background equations of motion and
the asymptotic expansions}\label{background}

\begin{equation}
\begin{split}
&0=f'+f \biggl(
\frac{r v_2'^2}{4v_2^2}+\frac{rv_1'^2}{2v_1^2} +\frac{rg'^2}{g^2}-\frac3r\biggr)
-\frac{s^2 r^3 (2 v_2^2+v_1^2)q^2}{8v_2 v_1^2}
-\frac{s^2 g^4(2 v_2^2 v_1^2+v_1^4+9)}{4v_2 v_1^2 r} 
\\&+\frac{2 g^2 s^2 (2 v_2+v_1)}{v_2 v_1 r}\,,
\end{split}
\eqlabel{eom1}
\end{equation}
\begin{equation}
\begin{split}
0=s'+\frac{s r}{4} \biggl(
\frac{v_2'^2}{v_2^2}+\frac{2 v_1'^2}{v_1^2}+\frac{4 g'^2}{g^2}
\biggr)\,,
\end{split}
\eqlabel{eom2}
\end{equation}
\begin{equation}
\begin{split}
&0=v_1''-\frac{v_1'^2}{v_1}+v_1' \biggl(
\frac{s^2g^4 (2 v_2^2 v_1^2+v_1^4+9)}{4f v_2 v_1^2 r}
-\frac{2 s^2 g^2(2 v_2+v_1)}{v_1 f v_2 r}
+\frac{s^2 r^3 q^2(2 v_2^2+v_1^2)}{8f v_2 v_1^2} +\frac 1r\biggr)
\\&-\frac{s^2g^4 (v_1^4-9)}{2v_1 f v_2 r^2}+\frac{s^2 r^2 v_2 q^2}{2v_1 f}
-\frac{4 s^2 g^2}{f r^2}\,,
\end{split}
\eqlabel{eom3}
\end{equation}
\begin{equation}
\begin{split}
&0=v_2''-\frac{v_2'^2}{v_2}+v_2' \biggl(
\frac{s^2g^4 (2 v_2^2 v_1^2+v_1^4+9)}{4f v_2 v_1^2 r}
-\frac{2 s^2g^2 (2 v_2+v_1)}{v_1 f v_2 r}
+\frac{s^2 r^3 q^2(2 v_2^2+v_1^2)}{8f v_2 v_1^2}+\frac1r\biggr)
\\&-\frac{s^2 (2 v_2^2 v_1^2-v_1^4-9)g^4}{2f v_1^2 r^2}
-\frac{4 s^2g^2}{f r^2} -\frac{s^2 q^2 r^2 (2 v_2^2-v_1^2)}{4f v_1^2}\,,
\end{split}
\eqlabel{eom4}
\end{equation}
\begin{equation}
\begin{split}
&0=g''-\frac{g'^2}{g}+g' \biggl(
\frac{g^4 s^2 (2 v_2^2 v_1^2+v_1^4+9)}{4f v_2 v_1^2 r}
-\frac{2 g^2s^2 (2 v_2+v_1)}{v_1 f v_2 r}
+\frac{s^2 r^3 q^2(2 v_2^2+v_1^2)}{8f v_2 v_1^2}+\frac1r\biggr)\\
&-\frac{s^2 g^5(2 v_2^2 v_1^2+v_1^4+9)}{2f v_2 v_1^2 r^2}
+\frac{2 g^3 s^2 (2 v_2+v_1)}{v_1 f v_2 r^2}\,.
\end{split}
\eqlabel{eom5}
\end{equation}

Eqs.~\eqref{eom1}-\eqref{eom5} should be solved numerically, subject to
the following asymptotic expansion
\nxt In the UV, \ie as $r\to 0$, and with the identification
$\ln [v_1v_2^{-1}]\Longleftrightarrow\calo_2$,
we have
\begin{equation}
\begin{split}
&f=1+f_3 r^3+\frac38 q^2 r^4-\frac16 v_{1,2} q^2 r^6+\calo(r^7)\,,\qquad
s=1-\frac32 v_{1,2}^2 r^4+\frac16 v_{1,2} q^2 r^6+\calo(r^7)\,,
\end{split}
\eqlabel{uuv1}
\end{equation}
\begin{equation}
\begin{split}
&v_1=1+v_{1,2} r^2+\biggl(
v_{1,4}+\left(\frac{24}{35} v_{1,2}^2-\frac{1}{35} q^2\right) \ln r\biggr) r^4
-\frac13 f_3 v_{1,2} r^5+\biggl(
v_{1,6}\\&+\biggl(
-\frac{13}{350} v_{1,2} q^2+\frac{156}{175} v_{1,2}^3\biggr) \ln r
\biggr) r^6+\calo(r^7\ln r)\,,
\end{split}
\eqlabel{uuv2}
\end{equation}
\begin{equation}
\begin{split}
&v_2=1-2 v_{1,2} r^2+\biggl(
\frac32 v_{1,2}^2+v_{1,4}+\frac18 q^2
+\biggl(
\frac{24}{35} v_{1,2}^2-\frac{1}{35} q^2\biggr) \ln r\biggr) r^4
+\frac23 f_3 v_{1,2} r^5
+\biggl(
v_{1,6}\\&-\frac{39}{10} v_{1,2} v_{1,4}+\frac{4647}{3500} v_{1,2}^3
-\frac{653}{3500} v_{1,2} q^2+\biggl(
\frac{13}{175} v_{1,2} q^2-\frac{312}{175} v_{1,2}^3\biggr) \ln r\biggr) r^6
+\calo(r^7\ln r )\,,
\end{split}
\eqlabel{uuv3}
\end{equation}
\begin{equation}
\begin{split}
&g=1+\biggl(-\frac{3}{56} v_{1,2}^2+\frac34 v_{1,4}
+\frac{1}{56} q^2+\biggl(
\frac{18}{35} v_{1,2}^2-\frac{3}{140} q^2\biggr) \ln r\biggr) r^4
+\biggl(-v_{1,6}+\frac{13}{10} v_{1,2} v_{1,4}\\&-\frac{1549}{3500} v_{1,2}^3
-\frac{37}{1750} v_{1,2} q^2\biggr) r^6+\calo(r^7\ln r)\,,
\end{split}
\eqlabel{uuv4}
\end{equation}
i.e.~the UV part of the solution is characterized (given $q$) by
\begin{equation}
\biggl\{\ f_3\,,\ v_{1,2}\,,\ v_{1,4}\,,\ v_{1,6}\biggr\}\,;
\eqlabel{uvpar}
\end{equation}
\nxt in the UV, \ie as $r\to 0$, and  instead with the identification
$\ln [v_1v_2^{-1}]\Longleftrightarrow\calo_1$,
we have
\begin{equation}
\begin{split}
&f=1+f_3 r^3+\frac38 q^2 r^4+\biggl(-\frac{9}{20}v_{1,1}^2f_3-\frac{3}{10}v_{1,1}q^2\biggr) r^5+\frac{37}{120}v_{1,1}^2 q^2r^6+\calo(r^7)\,,
\end{split}
\eqlabel{uva1}
\end{equation}
\begin{equation}
\begin{split}
&s=1-\frac34 v_{1,1}^2 r^2+\frac{489}{800} v_{1,1}^4 r^4
+\biggl(
v_{1,1}^5+\frac25 v_{1,1}^2 f_3+\frac{1}{10} v_{1,1} q^2\biggr) r^5
+\biggl(
\frac{5661}{22400} v_{1,1}^6\\
&+\frac18 v_{1,1}^3 f_3-\frac{269}{1680} v_{1,1}^2 q^2
+\frac34 v_{1,1}^2 v_{1,4}+\biggl(-\frac{51}{70} v_{1,1}^6
-\frac{3}{140} v_{1,1}^2 q^2\biggr) \ln r
\biggr) r^6+\calo(r^7\ln r)\,,
\end{split}
\eqlabel{uva2}
\end{equation}
\begin{equation}
\begin{split}
&v_1=1+v_{1,1} r-\frac15 v_{1,1}^2 r^2-\frac{31}{20} v_{1,1}^3 r^3
+\biggl(
v_{1,4}+\biggl(-\frac{34}{35} v_{1,1}^4-\frac{1}{35} q^2\biggr) \ln r\biggr) r^4
+\biggl(-\frac{103}{800} v_{1,1}^5\\
&+\frac{19}{60} v_{1,1}^2 f_3+\frac{11}{120} v_{1,1} q^2+\frac32 v_{1,1} v_{1,4}
+\biggl(
-\frac{3}{70} v_{1,1} q^2-\frac{51}{35} v_{1,1}^5\biggr) \ln r\biggr) r^5
+\biggl(
v_{1,6}\\&+\biggl(
-\frac{51}{70} v_{1,1}^6-\frac{3}{140} v_{1,1}^2 q^2\biggr) \ln r\biggr) r^6+\calo(r^7\ln r)\,,
\end{split}
\eqlabel{uva3}
\end{equation}
\begin{equation}
\begin{split}
&v_2=1-2 v_{1,1} r+\frac{13}{10} v_{1,1}^2 r^2+\frac{1}{10} v_{1,1}^3 r^3
+\biggl(
\frac{131}{40} v_{1,1}^4+\frac12 v_{1,1} f_3+v_{1,4}+\frac18 q^2
+\biggl(-\frac{34}{35} v_{1,1}^4\\&-\frac{1}{35} q^2\biggr) \ln r\biggr) r^4
+\biggl(
-\frac{4597}{400} v_{1,1}^5-\frac{14}{15} v_{1,1}^2 f_3-\frac{13}{30} v_{1,1} q^2
-3 v_{1,1} v_{1,4}+\biggl(
\frac{3}{35} v_{1,1} q^2\\&
+\frac{102}{35} v_{1,1}^5\biggr) \ln r\biggr) r^5
+\biggl(
\frac{166743}{14000} v_{1,1}^6-\frac{29}{40} v_{1,1}^3 f_3
+\frac{8061}{14000} v_{1,1}^2 q^2+\frac{39}{20} v_{1,1}^2 v_{1,4}+v_{1,6}
\\&+\biggl(-\frac{459}{175} v_{1,1}^6-\frac{27}{350} v_{1,1}^2 q^2\biggr) \ln r\biggr) r^6
+\calo(r^7\ln r)\,,
\end{split}
\eqlabel{uva4}
\end{equation}
\begin{equation}
\begin{split}
&g=1-\frac{3}{10} v_{1,1}^2 r^2-\frac12 v_{1,1}^3 r^3+\biggl(
\frac{2047}{1400} v_{1,1}^4+\frac18 v_{1,1} f_3+\frac34 v_{1,4}
+\frac{1}{56} q^2+\biggl(-\frac{51}{70} v_{1,1}^4\\&-\frac{3}{140} q^2\biggr) \ln r\biggr) r^4
+\biggl(
-\frac{73}{40} v_{1,1}^5+\frac{1}{10} v_{1,1}^2 f_3\biggr) r^5
+\biggl(
-\frac{6761}{14000} v_{1,1}^6+\frac{283}{240} v_{1,1}^3 f_3
\\&+\frac{2837}{21000} v_{1,1}^2 q^2
+\frac{19}{40} v_{1,1}^2 v_{1,4}-v_{1,6}+\biggl(
\frac{187}{700} v_{1,1}^6+\frac{11}{1400} v_{1,1}^2 q^2\biggr) \ln r\biggr) r^6+\calo(r^7\ln r)\,,
\end{split}
\eqlabel{uva5}
\end{equation}
characterized  (given $q$) by
\begin{equation}
\biggl\{\ v_{1,1}\,,\ f_3\,,\  v_{1,4}\,,\ v_{1,6}\biggr\}\,;
\eqlabel{uvpara}
\end{equation}
\nxt in the IR, \ie as $y\equiv 1-r\to 0$, we have
\begin{equation}
\begin{split}
&f=-\frac{(s^h_0)^2}{8v^h_{2,0} (v^h_{1,0})^2} \biggl(
2 (g^h_0)^4 \biggl((v^h_{1,0})^4+2 (v^h_{1,0})^2 (v^h_{2,0})^2+9\biggr)
-16 (g^h_0)^2 v^h_{1,0} \biggl(v^h_{1,0}+2 v^h_{2,0}\biggr)
\\&+q^2 \biggl((v^h_{1,0})^2+2 (v^h_{2,0})^2\biggr)
\biggr) y+\calo(y^2)\,,\\
&s=s^h_0+\calo(y)\,,\qquad v_i=v^h_{i,0}+\calo(y)\,,\qquad g=g^h_0+\calo(y)\,,
\end{split}
\eqlabel{ir}
\end{equation}
characterized  (given $q$) by
\begin{equation}
\biggl\{\ s^h_0\,,\ v_{1,0}^h\,,\ v_{2,0}^h\,,\ g^h_0\biggr\}\,.
\eqlabel{irpar}
\end{equation}

Given $q$, a numerical solution is
characterized by \eqref{uvpar} (or \eqref{uvpara}) and \eqref{irpar},
which determine the black membrane Hawking temperature $T$,
and the baryonic chemical potential $\mu_B$,
\begin{equation}
\begin{split}
&\frac {T}{|\alpha|}=\frac{s^h_0}{8 \pi v^h_{2,0} (v^h_{1,0})^2} \biggl(
2 (g^h_0)^2 \biggl(
8 v^h_{1,0} (v^h_{1,0}+2 v^h_{2,0})-(g^h_0)^2 ((v^h_{1,0})^4
+2 (v^h_{1,0})^2 (v^h_{2,0})^2+9)\biggr)
\\&-\biggl((v^h_{1,0})^2+2 (v^h_{2,0})^2\biggr)q^2\biggr)\,,
\end{split}
\eqlabel{tbaryonic}
\end{equation}
and, see \eqref{4dm},
\begin{equation}
\frac{\mu_B}{\alpha}=\frac1\alpha A^1_t\bigg|_{r=0}=-\int_0^1 \frac{qs}{v_2}\
dr\,.
\eqlabel{mubaryonic}
\end{equation}

\section{Equations of motion for $\zet_2$-odd fluctuations of the
baryonic black membranes}\label{flucm}

\begin{equation}
\begin{split}
&0=\cala_{2}'+\frac{c_2^2w}{c_1^2k} \cala_t'-\frac{2Fc_2^2w}{c_1^2k}\calv\,,
\end{split}
\eqlabel{flm1}
\end{equation}
\begin{equation}
\begin{split}
&0=\cala_t''+\biggl(-\frac{c_3'}{c_3}-\frac{c_1'}{c_1}+\frac{v_2'}{v_2}
+2 \frac{c_2'}{c_2}\biggr) \cala_t'-\frac{c_3^2 k}{c_2^2} (\cala_t k+\cala_{2} w)
-2 (F \calv)'+2 \calv F \biggl(
\frac{c_3'}{c_3}\\
&+\frac{c_1'}{c_1}-\frac{2c_2'}{c_2}-\frac{v_2'}{v_2}
\biggr)\,,
\end{split}
\eqlabel{flm2}
\end{equation}
\begin{equation}
\begin{split}
&0=\cala_{2}''+\biggl(
-\frac{c_3'}{c_3}+\frac{c_1'}{c_1}+\frac{v_2'}{v_2}\biggr)
\cala_{2}'+\frac{c_3^2 w}{c_1^2} (\cala_t k+\cala_{2} w)\,,
\end{split}
\eqlabel{flm3}
\end{equation}
\begin{equation}
\begin{split}
&0=\calv''+\biggl(
-\frac{c_3'}{c_3}+\frac{c_1'}{c_1}+\frac{2 c_2'}{c_2}
\biggr) \calv'
-\frac{2 F v_2}{c_1^2} \cala_t'+\biggl(\frac{2 v_2F^2}{c_1^2}
-\frac{c_3^2 (c_1^2 k^2-c_2^2 w^2)}{c_1^2 c_2^2}
\\&-\frac{8 g^2 c_3^2 (g^2 v_1 v_2-2)}{v_1}\biggr) \calv\,,
\end{split}
\eqlabel{flm4}
\end{equation}
\begin{equation}
\begin{split}
&0=\calb''+\biggl(
\frac{c_1'}{c_1}+\frac{2 c_2'}{c_2}-\frac{2 v_1'}{v_1}
-\frac{c_3'}{c_3}\biggr) \calb'
-\biggl(\frac{4 c_3^2 g^4 (v_1^2-3)}{v_2}
+\frac{c_3^2 (c_1^2 k^2-c_2^2 w^2)}{c_1^2 c_2^2}\biggr) \calb\,,
\end{split}
\eqlabel{flm5}
\end{equation}
where, compare with \eqref{4dm},
\begin{equation}
c_1=\frac{2\alpha \sqrt f}{r}\,,\qquad
c_2=\frac{2\alpha}{r}\,,\qquad c_3=\frac{s}{2r\sqrt f }\,,\qquad
F=\frac{q\alpha s}{v_2}\,.
\eqlabel{defcf}
\end{equation}
We explicitly verified that \eqref{flm1} is consistent with
\eqref{flm2}-\eqref{flm4}.

\bibliographystyle{JHEP}
\bibliography{membrane2v2}

\end{document}